\documentclass[aps,prl,twocolumn,showpacs]{revtex4}

\usepackage{amsmath}
\usepackage{amssymb}
\usepackage{graphicx}

\newcommand{\kb}{k_{\mathrm{B}}}
\newcommand{\nub}{\nu_{\mathrm{b}}}
\newcommand{\rhom}{\rho_{\mathrm{m}}}
\newcommand{\rhof}{\rho_{\mathrm{f}}}
\newcommand{\rhoc}{\rho_{\mathrm{c}}}
\newcommand{\ud}{\/\mathrm{d}\/}
\newcommand{\ple}{P^{(\mathrm{leq})}}
\newcommand{\hks}{h_{\mathrm{KS}}}

\begin{document}
\title{\bf Heat conduction and Fourier's law \\ by consecutive local mixing and thermalization}
\author{P. Gaspard}
\author{T. Gilbert}
\affiliation{Center for Nonlinear Phenomena and Complex Systems,\\
Universit\'e Libre de Bruxelles, Code Postal 231, Campus Plaine,
B-1050 Brussels, Belgium}
\begin{abstract}
We present a first-principles study of heat conduction in a class of models
which exhibit a new multi-step local thermalization mechanism which gives
rise to Fourier's law. Local thermalization in our models occurs as the
result of binary collisions among locally confined gas particles. We
explore the conditions under which relaxation to local equilibrium, which
involves no energy exchange, takes place on time scales shorter than
that of the binary collisions which induce local thermalization. The role
of this mechanism in multi-phase material systems such as aerogels is
discussed.
\end{abstract}

\maketitle

The transport property of thermal conductivity has been introduced as a
phenomenological law by Joseph Fourier in 1822 \cite{F1822}. Its molecular
origins were later discovered with Boltzmann's and Maxwell's works on the
kinetic theory of gases \cite{Boltz1896,Maxw1867}. In metallic solids, the
thermal conduction is mainly determined by the transport of electrons and
thus related to the electric conductivity according to the Wiedemann-Franz
law, as explained by Drude in 1900 \cite{Drude1900}. In contrast, the
thermal conductivity of non-metallic solids results from the mutual
scattering of sound waves due to the anharmonicities of the inter-atomic
forces, as shown by Peierls in 1929 \cite{Peierls1929,S06}. In all these
systems, the microscopic mechanism of thermal conduction is the scattering
of the energy carriers, which induces local thermalization in a
single-stage process, and subsequent uniformization of temperature
\cite{BLRB00}. Yet, this one-stage scenario does not exhaust the possible
microscopic mechanisms. Here, we prove that,  in a further class of
systems,  local mixing can precede the start of energy transfer leading to
local thermalization. On the basis of this mechanism, we provide a
derivation of Fourier's law. 

This new mechanism of local thermalization allows us to understand how
materials can become excellent thermal insulators, as is the case with
aerogels \cite{HP94}. In these materials, a gas is trapped in a solid-state
nanoporous matrix.  The gas molecules collide with the walls of the nanopores 
more frequently than among  themselves. Since the collisional transfer of
energy is smaller with a stiff wall than with another free molecule, the
changes of velocity orientation occur before a significant energy transfer.
Accordingly, we envisage the mechanism where the wall collision frequency
is shorter than the binary collision frequency.

In this respect, thermal conduction should proceed over three
well-separated time scales, $\tau_{\rm wall} \ll \tau_\mathrm{binary} \ll
\tau_\mathrm{macro},$ which are~: 
(i) the short time scale $\tau_\mathrm{wall}$ of the collisions on the solid
walls, inducing the mixing of the particle positions and velocity
orientations with 
negligible transfer of particle kinetic energy and no mass transport; 
(ii) the intermediate time scale $\tau_\mathrm{binary}$ of the binary
collisions between pairs of particles, 
achieving a local thermodynamic equilibrium at some
locally defined temperature; 
and (iii) the long time scale $\tau_\mathrm{macro}$ of the macroscopic
relaxation of the Fourier modes. 

To demonstrate this mechanism, we consider in this Letter Hamiltonian
dynamical systems in which hard-ball particles undergo elastic collisions 
either with immobile obstacles composing the solid matrix, or with each
other. The obstacles form a lattice structure of cells, each confining a
single mobile particle and preventing any mass transport. The motion of the
particles is controlled by the geometry of the cells in such a way that the
binary collisions only occur between nearest-neighboring mobile particles
at a rate which can be switched off by shrinking the domain of motion
within each cell.  At the critical geometry where the binary collisions
become impossible, the thermal conductivity vanishes with the binary
collision frequency, in a way which can be rigorously calculated, as we
show below.   The remarkable result is that Fourier's law of heat
transport can be established for  this class of systems,  based on the
strong ergodic properties that they enjoy \cite{BLPS92}.

Under the assumption that binary collisions are seldom compared to wall
collisions, the chaotic motion of individual particles within their cells
induces a rapid decay of statistical correlations. As a consequence, the
global multi-particle probability distribution of the system  typically
reaches local equilibrium distributions at the kinetic energy of each
individual particle before energy exchanges proceed. 
This allows us to
obtain a kinetic equation for the probability distribution of the  local
energies \cite{MMN84}, which, under a Boltzmann-type assumption,
reduces to a Boltzmann-Kac equation \cite{Kac56}. In contrast to
the Boltzmann equation which requires a further approximation, our kinetic
equation can be fully justified thanks 
to the mixing property of the dynamics and the separation of the time
scales  $\tau_\mathrm{wall} \ll \tau_{\rm binary}$.

\begin{figure}[htp]
  \centering
  \includegraphics[width = .44\textwidth]{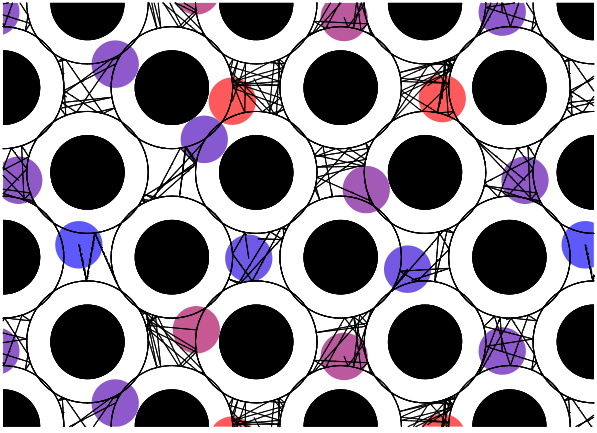}
  \caption{(Color online) 
    Example of lattice billiard with rhombic tiling. The mobile disks are
    color-coded from blue to red with growing kinetic energies. See text
    for details.}
  \label{fig.example}
\end{figure}

To be specific, we consider a system such as depicted in
Fig.~\ref{fig.example}. It consists of a periodic two-dimensional array of
discs of radii $\rhof$. In every cell between those discs, we place a single
mobile disc of radius $\rhom$, large enough that the particles are confined
to the regions between the discs. The shape of this region, which is a
bounded dispersing billiard \cite{CM06}, only depends on the sum of the two
radii $\rho = \rhof + \rhom$, which is kept constant. 

Energy exchanges occur when two moving particles
located in neighboring cells collide. Such binary collisions can take
place provided the radii
$\rho_\mathrm{m}$ of the moving particles is large
enough. For a lattice with cells of sides $l$, binary collisions are
thus possible above the critical value 
\begin{equation}
\rhoc \equiv \sqrt{\rho^2-(l/2)^2} < \rhom  < \rho \; .
\end{equation}
In this regime, 
the system is a thermal conductor. On the contrary, it becomes a 
thermal insulator for $\rhom \leq \rhoc$. The thermal conductivity vanishes
as $\rhom\stackrel{>}{\to}\rhoc$ in an exactly calculable way.   

In order to establish this result, we start from first principles.  The
system is of Hamiltonian character and the time evolution of the global 
probability distribution of the positions and velocities of all the $N$ 
particles is described by the pseudo-Liouville equation \cite{EDHVL69}~: 
\begin{eqnarray}
\lefteqn{\partial_t \, p_N =}\label{pseudoliouville}\\ 
&&\sum_{a = 1}^N \left[
-\mathbf{v}_a \cdot \partial_{\mathbf{r}_a}
+ \sum_{k=1}^d K^{(a,k)}\right] p_N
+ \frac{1}{2}\sum_{a,b=1}^N B^{(a,b)} p_N \, .
\nonumber
\end{eqnarray}
where $-\mathbf{v}_a \cdot \partial_{\mathbf{r}_a}$ is the advection term
expressing the changes of the probability distribution due to the free
motion at the velocity $\mathbf{v}_a$ of particle $a$
inside its cell, the wall term $K^{(a,k)}$ is
the operator ruling the collisions of particle $a$ on the fixed disc $k$
in the corresponding cell,
and $B^{(a,b)}$ rules the binary collisions between the particles $a$ and
$b$. Without these last terms, the pseudo-Liouville equation would
describe the relaxation of the probability distribution to the product of
local distributions $\prod_{a=1}^N \delta(\epsilon_a -m v_a^2/2)$
for all the particles at their respective kinetic energies $\epsilon_a$.
This relaxation is driven by the local mixing of the single-cell
dispersing billiards, which takes place over
the short time scale $\tau_\mathrm{wall}$.
The important point is that the
probability distribution has not yet reached a local Maxwellian
distribution.  The local thermalization is the outcome of the binary
collisions which take place on the intermediate time scale $\tau_{\rm
  binary}$. This stage of the process of heat conduction is in general
utterly difficult to master in a mathematically controllable way.  However,
the fact that local mixing precedes local thermalization allows us to
obtain a kinetic equation for the probability distribution of the energies
of the particles~: 
\begin{eqnarray}
\lefteqn{\ple_N(\epsilon_1, \dots, \epsilon_N, t) \equiv
\int \prod_{a=1}^N\ud \mathbf{r}_a\, \ud \mathbf{v}_a\times}\label{ple}
\\
&&p_N(\mathbf{r}_1,\mathbf{v}_1,\dots,\mathbf{r}_N,\mathbf{v}_N,t)
\prod_{a=1}^N \delta(\epsilon_a -m v_a^2/2),
\nonumber
\end{eqnarray}
where $v_a \equiv \Vert\mathbf{v}_a\Vert$. 
Integrating the pseudo-Liouville equation (\ref{pseudoliouville}) with
the local equilibrium distributions, the advective terms as well as the
terms of the wall collisions vanish.
We remain with the terms due to the binary
collisions, which yield the following master
equation \cite{MMN84}~:
\begin{eqnarray}
\lefteqn{\partial_t \ple_N(\epsilon_1, \dots, \epsilon_N, t) = 
\frac{1}{2}\sum_{a,b = 1}^N \int d\eta \times}
\label{masterequation}\\
&&\Big[ W_{(\epsilon_a + \eta, \epsilon_b - \eta)\to(\epsilon_a,
\epsilon_b)} \;
\ple_N(\dots, \epsilon_a + \eta, \epsilon_b - \eta, \dots, t)
\nonumber\\
&&- W_{(\epsilon_a, \epsilon_b)\to(\epsilon_a - \eta, \epsilon_b + \eta)}
\; \ple_N(\dots, \epsilon_a, \epsilon_b, \dots, t)\Big] ,
\nonumber
\end{eqnarray}
where 
\begin{eqnarray}
  \lefteqn{W_{(\epsilon_a, \epsilon_b)\to(\epsilon_a - \eta, \epsilon_b +
      \eta)} =
    \frac{2\rhom m^2}{(2\pi)^2|\mathcal{L}_{\rho,\rhom}(2)|}\times}
  \label{energytransition}\\
  &&\int\ud\phi\, \ud \mathbf{R} 
  \int_{\hat{\mathbf{e}}_{ab}\cdot\mathbf{v}_{ab} > 0}
  \ud \mathbf{v}_a \, \ud \mathbf{v}_b
  \, \hat{\mathbf{e}}_{ab}\cdot\mathbf{v}_{ab}
  \, \delta\left(\epsilon_a - \frac{m}{2}v_a^2\right)
  \times\nonumber\\
  &&\delta\left(\epsilon_b - \frac{m}{2}v_b^2\right)
  \delta\left(\eta -
    \frac{m}{2}[(\hat{\mathbf{e}}_{ab}\cdot\mathbf{v}_a)^2  -  
    (\hat{\mathbf{e}}_{ab}\cdot\mathbf{v}_b)^2 ]\right),
  \nonumber
\end{eqnarray}
is the transition rate of the binary collisions of energy transfer $\eta$
between the neighboring particles $a$ and $b$. In this expression,
$|\mathcal{L}_{\rho,\rhom}(2)|$ is the volume 
of the billiard corresponding to two neighboring cells $a$
and $b$, which can be approximated by the square of the 
volume of a single cell $|\mathcal{B}_\rho|^2$,
$\phi$ is the angle of the unit vector $\hat{\mathbf{e}}_{ab} =
(\cos\phi,\sin\phi)$ connecting $a$ and $b$, and $\mathbf{R} = 
(\mathbf{r}_a + \mathbf{r}_b)/2$ the center of mass of the particles $a$
and $b$. The transition rate $W$ can be further expressed
in terms of Jacobian elliptic functions \cite{GG08}. The 
master equation (\ref{masterequation}) describes the time
evolution as a continuous-time stochastic process of Poisson type with
exponential probability distributions for the waiting times between the
random events of energy transfers due to the binary collisions.   
Accordingly, the master
equation (\ref{masterequation}) shows that local
thermalization is reached over the intermediate time scale $\tau_{\rm
  binary}\simeq 1/\nub$ of the binary collisions, 
which, in terms of the binary collision frequency, is given by
\begin{eqnarray}
  \nub &=& \beta^2 \int \ud\epsilon_a \, \ud\epsilon_b \, \ud \eta \, 
  W_{(\epsilon_a, \epsilon_b)\to(\epsilon_a - \eta, \epsilon_b + \eta)}
  \, e^{-\beta(\epsilon_a+\epsilon_b)},\nonumber\\
  &=& \sqrt{\frac{\kb T}{\pi m}}\frac{2\rhom}{|\mathcal{B}_\rho|^2}
  \left(\int\ud\phi\,\ud\mathbf{R}\right)  ,
  \label{nubKT}
\end{eqnarray}
where $\beta=1/(\kb T)$ is the inverse of the temperature $T$
multiplied by Boltzmann's constant $\kb$. 

Starting from the master equation (\ref{masterequation}), we derive the
heat equation 
\begin{equation}
\partial_t T= \partial_x (\kappa \, \partial_x T),
\label{fourierlaw}
\end{equation}
for the local temperature $T=T_a$ defined in terms of the kinetic energy of
particle $a$ averaged over the probability distribution (\ref{ple}),
$\langle \epsilon_a \rangle \equiv \kb T_a$. 
The derivation of Eq.~(\ref{fourierlaw}) from stochastic models such
as defined by the master equation (\ref{masterequation}) requires the
computation of correlation functions \cite{Spohn91}. 
In particular the heat current between two neighboring cells involves an
average with respect to the joint probability distribution of their
energies and thus depends on their correlations. Assuming non-equilibrium
boundary conditions, a systematic computation of the stationary
state can be performed through a cluster expansion. Single cell
distributions, which solve the Boltzmann-Kac equation \cite{Kac56}, give
the first order solutions of this expansion, 
while pair correlations appear at the second order, and so on. 
For the stochastic system described by Eq.~(\ref{masterequation}), it turns
out that pair correlations are of second degree  in the local temperature
gradients and can therefore be neglected in the computation of the transport
coefficient, which involves only terms linear in this quantity. We can
thus limit the cluster expansion to first order 
and find the expression of the coefficient of thermal conductivity~:   
\begin{equation}
\frac{\kappa}{l^2} 
= \sqrt{\frac{\kb T}{\pi
     m}}\frac{2\rhom}{|\mathcal{B}_\rho|^2} 
\left(\int\ud\phi\, \ud\mathbf{R}\right).
\label{heatconductivity}
\end{equation}
We verified this result by direct numerical simulations of
Eq.~(\ref{masterequation}). Alternatively, $\kappa$ can be defined through
a Green-Kubo formula. The result (\ref{heatconductivity}) shows that
$\kappa$ is given by the $\delta(t)$ part of the 
energy current-current correlation discussed in Ref. \cite{Spohn91}. We
notice that this result holds here even though the heat current does not
have the gradient form of \cite{Spohn91}, {\em i.~e.} it cannot be written
as the difference of two local functions. The details will be
presented elsewhere  \cite{GG08b}.  

The comparison between Eqs. (\ref{nubKT}) and (\ref{heatconductivity})
shows the equality of the thermal conductivity and collision frequency
of the stochastic process described by Eq.~(\ref{masterequation}). 
By extension, the same holds of the corresponding properties of 
the billiard system in the limit $\rhom\to\rhoc$~: 
$\lim_{\rhom\to\rhoc} \kappa/(l^2 \nub) = 1.$
Furthermore, the thermal conductivity can be explicitly calculated in this
limit and shown to vanish as the third power of the difference
$\rhom-\rhoc$.  Indeed, the geometric integral appearing in both 
Eqs. (\ref{nubKT}) and (\ref{heatconductivity})
vanishes as
\begin{equation}
 \int\ud\phi\, \ud\mathbf{R} =  c_3 (\rhom-\rhoc)^3 + c_4 (\rhom-\rhoc)^4 +
 \cdots 
 \label{intphiR}
\end{equation}
where the coefficients in this expansion can be computed 
according to the specific geometry of collision events \cite{GG08}.

\begin{figure}[htb]
  \centering
  \includegraphics[width = .45\textwidth]{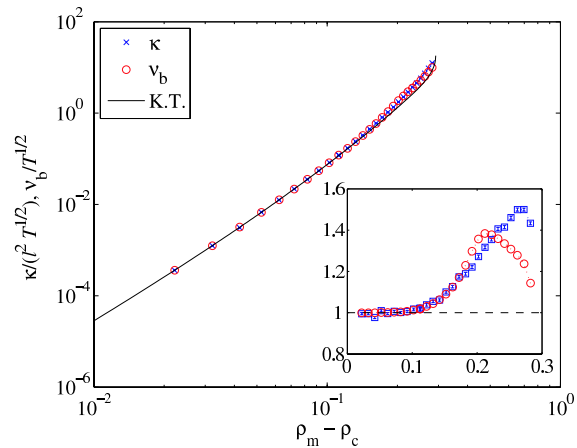}
  \caption{(Color online) Thermal conductivity $\kappa$ in reduced units,
    computed by molecular dynamics from the heat exchange of a single
    particle with stochastically thermalized neighbors,  and binary
    collision frequency $\nub$, as  
    functions of $\rhom - \rhoc$, compared to our theoretical calculations 
    (\ref{nubKT}) and (\ref{heatconductivity}) (solid line). 
    The inset shows the ratios between $\nub$ and $\kappa$ and the RHS of 
    (\ref{nubKT}) and  (\ref{heatconductivity}). The computations were
    performed with a rhombic cell in contact with stochastically thermalized
    cells on each sides, in which case
   $c_3 = 128 \rhoc/(3 l^2)$ and $c_4 = 256 \rhoc^2/(3 l^4)$
    in Eq. (\ref{intphiR}). The sum of radii is here fixed to the value
    $\rho =\rhof+\rhom= 9/25$.} 
  \label{fig.kappa}
\end{figure}

Numerical computations of the quantities computed in Eqs.
(\ref{nubKT}) and (\ref{heatconductivity}) are shown in
Fig.~\ref{fig.kappa}, asserting the validity of the stochastic description
(\ref{masterequation}) on the one hand and the computation of its
properties on the other.

Finally, the local mixing property can be characterized by the Lyapunov
exponents measuring the stretching and contraction of phase-space volumes.
Remarkably, all these exponents can be analytically calculated in the same
critical limit where the binary collision frequency vanishes. In this
limit, the particle velocities share a Maxwellian probability distribution
at the inverse temperature $\beta$ while the particles become
independent. Accordingly, a system containing $N$ particles presents the
$N$ positive Lyapunov exponents  
\begin{equation}
\lambda_i = \lambda_+ \sqrt{\frac{2}{m\beta}\ln\frac{N}{i-1/2}},
\quad i = 1,\dots,N,
\label{lambdai}
\end{equation}
where $\lambda_+$ is the positive Lyapunov exponent of a single isolated
particle moving at unit speed. $2N$ exponents are zero in the said limit
and the $N$  remaining exponents take the negative values $-\lambda_i$. The
sum of all the exponents vanishes since the system is Hamiltonian and
satisfies Liouville's theorem. The system is thus chaotic with
an extensive Kolmogorov-Sinai entropy per unit time given by $\hks = N
\lambda_+ \sqrt{\pi/(2m\beta)}$. 

\begin{figure}[htb]
  \centering
  \includegraphics[width = .45\textwidth]{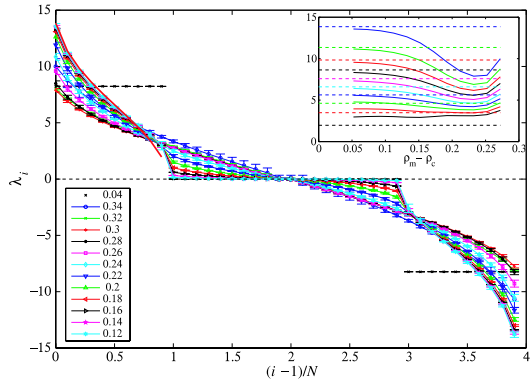}
  \caption{(Color online) Lyapunov exponents $\lambda_i$ versus their index
    $i$. The numbers in the legend correspond to the values of $\rhom$. The
    first one, $\rhom = 0.04 < \rhoc$, yields the positive exponent
    $\lambda_+$ associated to isolated cells, with all the particles at
    the same speed. The solid red line is the first half of the
    spectrum as predicted by Eq.~(\ref{lambdai}). The inset shows
    $\lambda_i$ versus $\rhom-\rhoc$ for the first half of the positive
    exponents, $\lambda_1,\dots,\lambda_N$ and compares them to the
    asymptotic value (\ref{lambdai}) (dashed lines). The computations were
    performed with a rhombic channel of size $N=10$ cells, $\rho = 9/25$.} 
  \label{fig.lyap}
\end{figure}

Figure \ref{fig.lyap} shows the agreement between the numerically computed
Lyapunov exponents and Eq. (\ref{lambdai}), in the limit where the system
becomes insulating. 

To conclude, we have presented in this letter
a mechanism of heat conduction which proceeds
in two stages and involves no
mass transport.  The first stage is a local mixing 
phase which takes place in the individual cells and involves no  energy
transfer between them. This stage thus induces the relaxation of phase
space distributions to local equilibrium without thermalization.
The second stage is the energy transfer 
which leads to local thermalization
and can be efficiently described in probabilistic terms by means of our
master equation.  
The uniformization of the temperature over the whole system can then
proceed as described by the heat equation. 
Whereas the first-principles derivation of Fourier's law is 
problematic based on the standard one-stage local thermalization mechanism,
it is remarkable that it can be carried out in the presence of this
additional local equilibrium mechanism.
We notice that both mechanisms correspond to 
physically distinct limits for the heat conductivity.   
In the standard one-stage mechanism, the ballistic transport of the energy
carriers is hampered by scattering so that the theory should there
demonstrate a finite conductivity with respect to an unperturbed situation
where the conductivity is unbounded.  In contrast, in the two-stage
mechanism we here describe, the thermal conductivity vanishes in the
reference system, allowing the first-principles derivation of Fourier's law.
This two-stage mechanism is encountered in aerogels where the thermal
conductivity takes the smallest known values.  Our analysis therefore
suggests that fundamental insight into the phenomenon of heat conductivity
could be obtained by the experimental study of aerogels. 

\begin{acknowledgments}
The authors wish to thank D. Alonso, J. Bricmont, J. R. Dorfman, A. Kupiainen,
R. Lefevere, C. Liverani, C. Mej\'{\i}a-Monasterio and S. Olla  for
fruitful discussions and comments. This research is financially supported
by the Belgian Federal 
Government  (IAP project ``NOSY") and the ``Communaut\'e fran\c caise de
Belgique'' (contract ``Actions de Recherche Concert\'ees''
No. 04/09-312). TG is financially supported by the Fonds de la Recherche
Scientifique F.R.S.-FNRS.
\end{acknowledgments}


\end{document}